# Unraveling the Interplay of Leaf Structure and Wettability: A Comparative Study on Superhydrophobic Leaves of Cassia tora, Adiantum capillus-veneris, and Bauhinia variegata

*Shubham S. Ganar*[1] *and Arindam Das*[1]*

[1]School of Mechanical Sciences, Indian Institute of Technology (IIT) Goa, GEC Campus, Farmagudi, Ponda, Goa 403401, India

KEYWORDS

Surface morphology, Wettability, Superhydrophobic leaves, Drop impact

ABSTRACT

In this article, superhydrophobic leaves of Cassia tora, Adiantum capillus-veneris (ACV), and Bauhinia Variegate are reported for the first time, and the wettability of these leaf's surfaces was correlated with their surface morphology at micro and nanoscale. Field Emission Scanning Electron Microscopy (FESEM) images of the surfaces were used to get surface morphological information at the micro-nanoscale structures. A special drying method was implemented to ensure the minimal structural collapse of these surfaces under the high vacuum of FESEM. FESEM images of Cassia tora leaves showed widely spaced, low aspect ratio nano petals distributed on bumpy blunt micro features, responsible for high contact angle hysteresis and high roll angle measured on the Cassia tora leaves. ACV leaves showed the presence of micron-scale spherical morphology made of nanoscale hair-like features. These hierarchical re-entrant surface features generated a very high contact angle and low roll-off angle. Leaves of Bauhinia variegate showed similar superhydrophobic and self-cleaning properties. However, surface features were different, which consisted of a higher aspect ratio and closely spaced nano petals uniformly distributed over flat surfaces consisting of micro-scale ridges. Droplet impact studies on these surfaces at different Weber numbers showed different behaviour due to these different micro-nano features.

## 1. Introduction

Nature has a wealth of functional surfaces that have evolved over millennia. Plants and animals have developed surfaces with varying degrees of wettability, ranging from extremely slippery to highly sticky to extremely water-repellent surfaces with a contact angle for water (CA) greater than 150º, with a low roll-off angle of less than 10º, are known as superhydrophobic self-cleaning surfaces (SHSs)[1] such as leaves Nelumbo nucifera[2]. This effect is widely known as "the lotus effect". Numerous other natural surfaces have superhydrophobicity, such as taro leaves[3], rose petals[4], namib desert beetle[5], and water strider[6] etc. Combination of micro-nano scale structures[7], along and low-surface energy materials[8], are essential to generate superhydrophobicity. The superhydrophobic leaf surfaces have hierarchical morphology where nanoscale features made of hydrophobic wax of low surface energy are distributed over microscale features. Inspired by these natural surfaces researchers and engineers have developed many new materials and technologies, mimicking these surfaces. These artificial surfaces have shown tremendous potential in many applications in healthcare, transportation, and energy and many other fields. Synthetic superhydrophobic surfaces found application anti-icing, self-cleaning, drag reduction, and antifouling surfaces, etc[9–13]. Hence it is important to study different superhydrophobic leaves which may show new surface architecture and facilitate more alternate solution for practical problems through biomimicking.

In an extensive study, Neinhuis and Barthlott[2] reported 200 water-repellent plant leaf surfaces and their wetting properties. It has been found that the epidermis papillose (microstructure) and various shapes of wax crystals on the surface are responsible for superhydrophobicity on plants' leaves and petals. They reported that these wax crystals create nano roughness on top of microfeatures[14]. Wheat leaves showed a superhydrophobicity because of micro/nano-scale roughness; the nano features on the leaves are made of hydrophobic wax[15]. The chemical composition of wheat wax includes primary alcohol, esters, alkanes, fatty acids, and beta-amyrines[16]. Apart from the lotus leaves, many researchers found water-repellent leaves like Taro leaves[3], Clover leaves[17,18], Rice leaves[19], and Reed leaves[20]. In a recent study, Fritch et al.[21] identified and reported static and dynamic wettability of the three New Zealand native plants: Arthropodium bifurcatum, Euphorbia glauca, and Veronica Albicans. From earlier plant studies, it is evident that enhanced non-wettability properties arise from the irregularity or structure and shape of both epicuticle wax and epidermal cells, respectively[3,8,16,19,22,23].

Researchers have drawn inspiration from the non-wetting properties of several plant leaves to build such practical non-wetting surfaces. Inspired by the lotus leaf, researchers have developed a water-repellant surface with a hierarchical structure and also studied the static and dynamic wettability of the surface[8]. They have observed that water on the surface goes into the shift from the Cassie–Baxter to Wenzel state with an increase in the spacing between the two consecutive pillars in static conditions[8]. Bharat Bhushan et al.[8] explained the dynamic behaviour of water droplets on silicon surfaces with pillars micropatterns of two different diameters and heights with varying pitch values. They found that dynamic effect, i.e., bouncing of a droplet, can destroy the composite solid-air-liquid interface on the textured surface. Wettability measurements on artificial honeycomb geometry surfaces bioinspired from taro leaf[3] showed geometric parameters strongly influence static and dynamic wettability. Patil et al.[24] investigated the transition between sticking, partial bouncing, and complete rebound and subsequently related experimental observation to the impact velocity on different micropillar surfaces. Amid the different roughness structures reported for superhydrophobicity, the re-entrant microstructures provide excellent stability in the Cassie-Baxter wetting state. Microstructures such as micro-hoodoo, T-shaped pillars, spheres, micro-mushrooms etc., have re-entrant morphologies[25]. It creates an extra energy barrier for transitioning between the Cassie-Baxter and Wenzel wetting state. That stabilizes the heterogeneous air-liquid interface and prevents the intrusion of the liquid into the roughness features compared to non-re-entrant structures[26]. Artificial surfaces with re-entrant structures have already shown extreme non-wetting properties that can support the Cassie state with water and different organic liquids[27,28].

The current study aimed to examine the unique topographical features and wetting properties of such three superhydrophobic leaves of three plants, Cassia tora, Adiantum capillas veneris, and Bauhinia variegata, showing very high-water contact angle > 150º (see Figure 1). These leaves were collected from respective plants found in the IIT Goa campus. To the best of our knowledge, neither the wettability nor the surface morphology of these plant leaves was reported in the literature. In the current report, wettability studies on these surfaces and their correlation with their micro-nano features were reported for the first time. A low-cost FESEM sample preparation method was developed to ensure minimal surface damage and collapse of micro nano features due to the dehydration process. The surface wettability of the plant leaves was assessed by measuring the static contact angle, contact angle hysteresis (CAH), and roll-off angle. Drop impact tests were performed on these surfaces to observe the dynamic wettability. The correlation between the wettability and micro-nano features on the surface was established through suitable theoretical and data analysis. This study thoroughly explains how the surface features present on these leaves affect dynamic and static wettability.

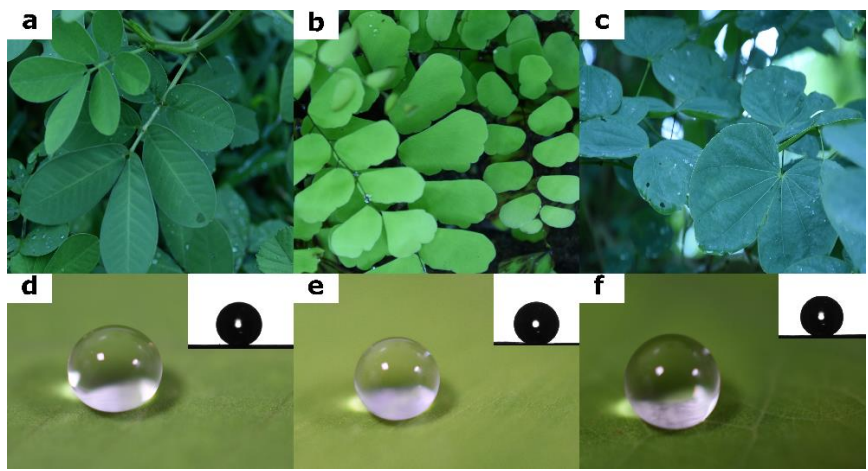

**Figure 1.** Cassia tora (a & d), Adiantum capillus-veneris (ACV) (b & e), and Bauhinia variegata (c & f) with water droplets on the corresponding leaves and goniometric images, respectively.

## 2. Experimental Section

### 2.1 Sample Preparation & Surface Characterization

Fully grown leaves from three plants were collected just before the measurements to ensure that the leaves were fresh and healthy during the experiment. (If required, store them in a highly humid chamber to prevent crumpling or distortion of a surface morphology). The leaves were gently cleaned with deionised water and nitrogen gas to remove any dust particles from the surface. FESEM (Carl Zeiss-Sigma 300) was used to capture surface morphology at the micro-nano scale. Under the high vacuum environment of FESEM, epidermal cells, wax microstructure, and trichomes are expected to collapse structurally. To prevent such deformations, the leaves must be dehydrated. Many methods were reported to dehydrate the sample, such as critical point drying (CPD), freeze-drying (lyophilising), and chemical fixation, which are expensive and complicated[29–32]. In the current report, a simple non-chemical air-drying method was used for the sample preparation. The leaves samples were allowed to dry slowly at room temperature of 24°C in a vacuum desiccator kept at internal pressure a little above water saturation pressure (10% above the saturation pressure) for 36 hrs. Samples prepared through this protocol didn't show any sign of structural collapse except at very few locations where few shrinkages of the epidermal cells were observed. This sample preparation method seems to be better than other methods where significant structural collapses were observed[33]. These dehydrated samples were cut, mounted on the stubs, and coated with 3nm thick conductive gold layer before being placed inside the FESEM chamber to avoid charge build-up. To have a superhydrophobic surface through biomimicking from the plant leaves, studying the chemical configurations of the wax-like constituents containing the leaves surfaces is essential. In addition to their surface structures, chemical configurations present on the surface regulate the surface free energy. Thus, chemical composition significantly affects wettability, regardless of surface morphology/roughness[34]. Hence FTIR spectra for Cassia tora, Adiantum capillus-veneris (ACV) and Bauhinia variegata have been measured and analyzed.

**2.2 Wettability Measurements**

Fresh test leaves were mounted on the goniometer (Rame Hart, Model 500) stage to measure Young's contact angle, advancing-receding contact angles and droplet roll-off angles. Contact angles were measured by vertically placing sessile DI water drops of the fixed volume of 5µl droplets on test surfaces. For each type of leaf, a total of ten measurements were taken (five different locations of each leaf on two different leaves of the same plant). For all experiments, the temperature and relative humidity of the surroundings was maintained at $24^0C$ and 75%, respectively. A goniometer and a monochrome video camera were used to capture the images of drops. The drop volume-changing method was used to measure dynamic contact angles and contact angle hysteresis (CAH). In this method, the needle was brought close to the leaf surface, and subsequently, the water was pumped gradually to raise the drop volume. The advancing contact angle was determined just before the sudden movement of the three-phase contact line (TPCL) in the advancing direction. Similarly, the receding contact angle was measured during the suction of the droplet when TPCL was about to retract[35]. The difference among the advancing and receding angles gave the value of CAH. On the other hand, droplet roll-off angles were measured by keeping sessile water on the test surface and tilting it (tilting the goniometer sample stage) slowly till the droplet rolled off from the surface. The angle at which drops start rolling over the surface is known as the droplet roll-off angle.

**2.3 Droplet Impact Study**

Droplet impact tests were performed on leaves by placing them on smooth glass slides and clamping them from both ends. The droplets of diameter 2.8 mm were created at the tip of the Teflon-coated needle connected to a syringe mounted on the Harvard syringe pump operating with an infuse rate of 1 ml/hr. The impact velocity $V_i$ of the droplets is varied simply by changing the fall height from 10 mm to 150 mm, which corresponds to the impact of 0.48 ms−1 to 1.7 ms−1. Corresponding Weber number We value for these experiments were 9,30,62 is 105. The Weber number is the ratio between inertia and surface tension forces, defined as $We=(\rho D V_i^2)/\sigma$, with $\sigma$ and $\rho$ being the surface tension and density of water respectively, where D is the droplet diameter. The dynamics of droplet impact dynamics were captured from side view, using Phantom V4.2 high-speed camera at 1024 X 512 pixels resolution and 5000 frames per second. A high-beam light source is placed behind the substrate such that the light source, substrate, and high-speed camera are on the same optical axis, as shown in Figure 2. The video analysis was performed using MATLAB, and ImageJ software was used to extract all required information from the images belonging to different regimes of drop impact.

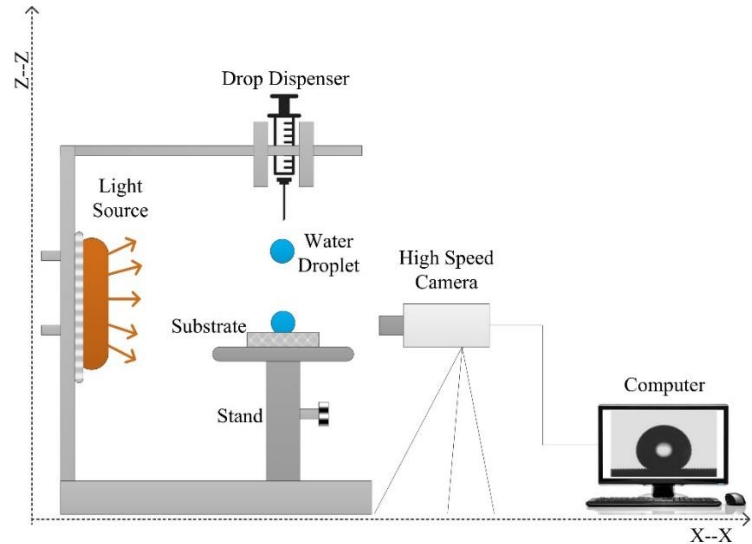

**Figure 2.** Schematic diagram of the experimental setup.

## 3. Results and Discussions

### 3.1 Analysis of Surface Topography

Air-dried leaves were observed inside the FESEM, and corresponding images were taken at four different magnifications and shown in different rows of Figure 2 for all three plants (column-wise). There is an apparent variation between the microstructures and the nano features for different specimens.

Cassia Tora: Cassia tora, also known as senna tora, is a plant species in the family Fabaceae, and subfamily Caesalpinioideae are primarily found in the central and western regions of the Indian subcontinent. It grows wildly and is consumed as a green leafy vegetable in the monsoon. FESEM micrographs of the Cassia tora leaf (topside) are shown in Figure 3. (a-d). The leaf was found to be completely covered with a pattern of large asymmetrical papillae or micro-bumps with typical sizes of 40μm - 60μm, as shown in Figure 3. (a). A valley surrounds each bump and keeps them disconnected from other bumps. The micro-bumps on these leaves are covered with irregular crenate platelets (ICP)[36] or a layer of epicuticular wax platelets, each featuring sharp edge nano features that are sparsely distributed over the surface. This unique combination of micro and nanostructures is what gives these leaves their superhydrophobic properties. The edges of these ICP overlapped the adjacent platelets and contributed toward the complex nanomorphology. Our observations of the epicuticular wax platelets of Cassia tora were similar to those described in earlier reports for T. aestivum[16,37].

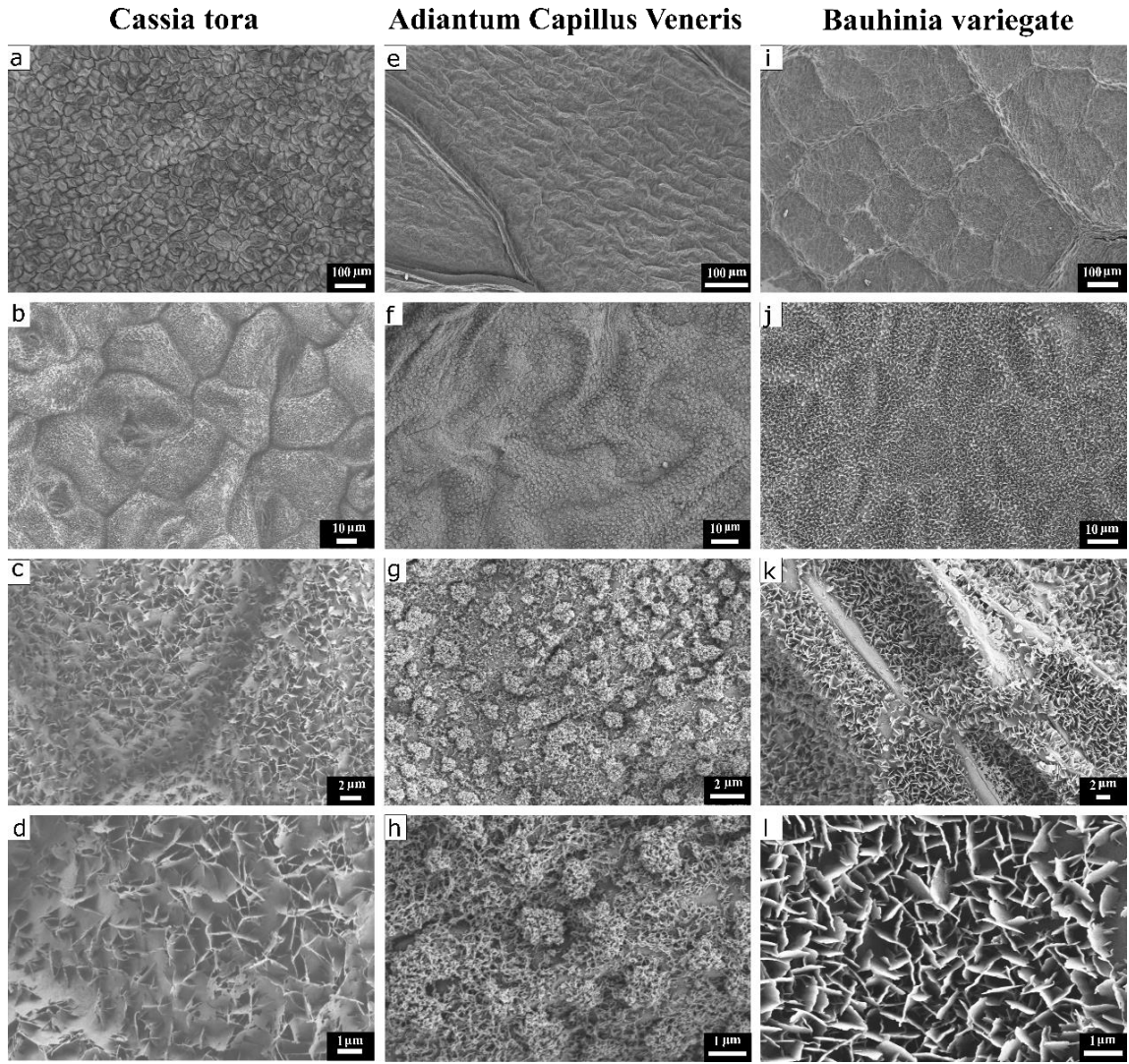

**Figure 3.** SEM of for Cassia tora (a-d), Adiantum capillus-veneris (ACV) (e-h), and Bauhinia variegate (i-l) leaf, respectively.

Adiantum capillus-veneris (ACV): It is a fern species in the genus of adiantum and belongs to the family pteridaceae. The Adiantum comes from the Greek word "Adiantos", which means un-wettable, and the remaining word capillus veneris means the hair of Venus, goddess of love. It is cultivated as a popular garden plant and is mainly found in temperate and tropical regions with high moisture content but not saturating [38]. Figure 3. (e-h) shows the FESEM images of ACV; the leaf surface is entirely different from both leaves. The leaf surface has a hemispherical shape papillae micro-feature with the characteristic dimensions around 0.5μm - 2μm as shown in Figure 3. (g-h). This micro-feature is evenly spread over the entire surface with spacings in the range of 0.5μm – 4μm between them. The zoomed FESEM images of the leaf (See Figure 3. g-h) showed that these hemispherical micro-textures were a grouping of hair-like nano features. These nano features overlap each other to form porous hemispherical shapes. The leaf surface in between these micro balls of nano hair structures was also covered with a thin layer of branchlike nano features, i.e., a membranous extension of the crystalloid (membranous platelet epicuticular wax)[36] as shown in Figure 3. (h). Thus, ACV leaves are expected to be highly hydrophobic due to the intricate hierarchical architecture of the papilla and three-dimensional re-entrant wax projections.

Bauhinia variegate (BV): It is a fast-growing tree found in humid regions of the world. Fabaceae is one of the most found flowering plants in southeast Asia and the Indian subcontinent. The leaves are long and broad, obcordate in form, and rounded at the base and apex, measuring 10–20 cm[39]. The morphology of the Bauhinia variegate has a flat surface having a network of micro ridges dividing the surface into many small regions of 80 μm to 200 μm circular or polygonal shapes. The region within the micro ridges has variable elliptical geometry, as shown in Figure 3. (i). These micro ridges have characteristic dimensions around 10μm to 20μm. High-resolution FESEM images show the presence of sharp petal-like nanofeatures epicuticular wax crystolloid[36,40] on top of the entire leaf surface, including the micro ridges (Figure 3.(j-l)). These nano features are closely packed, sharper and have a higher aspect ratio when compared to the nanostructures on the surface of Cassia tora leaf. A similar structure can be found on Colocasia esculenta leaves, which make the leaf superhydrophobic[41]. Since micro ridges and nano texture increases the roughness factor[8] and facilitate air pockets formation at the water-solid interface, thus contributing to the hydrophobic property of the plant surface[7]. Because of this densely packed hierarchical structure, the Bauhinia variegate leaves showed superior superhydrophobicity.

Fourier Transform Infrared (FTIR) spectroscopy was conducted and analyzed to obtain spectra,(as shown in Figure 4) to gain insights into the chemical compositions covering these plant leaves. Remarkably, the FTIR spectra of Cassia tora, Adiantum capillus-veneris (ACV), and Bauhinia variegata leaves exhibit a surprising similarity, suggesting that the chemical constituents present on these leaves might exist identical or very similar. The thick and intense band ranging from 3200 to 3600 cm$^{-1}$ is due to OH groups influenced by hydrogen bonding. While the peak at 1029.5 cm$^{-1}$ is attributed to alcoholic C-O stretching. A C=O extending from the carboxylate group can be assigned to the strong peak at 1643.3 cm$^{-1}$. Furthermore, the band at 2918.7 to 2850 cm$^{-1}$ is caused by CH spreading vibrations of the CH, CH2, and CH3 groups[34].

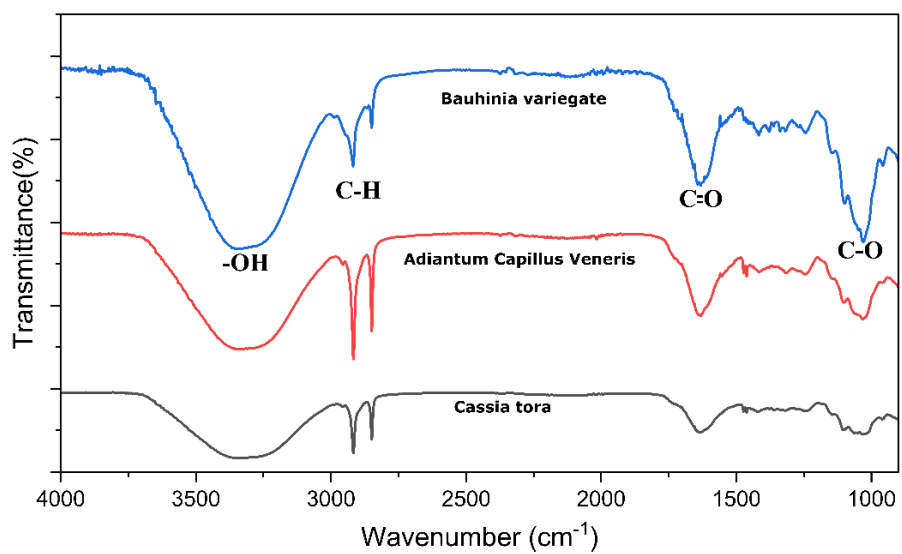

**Figure 4.** Fourier transform infrared (FTIR) spectra analysis of different leaves.

## 3.2 Surface wettability

The static wettability and dynamic wettability of leaves were evaluated by measuring the equilibrium contact angles and dynamic contact angles (advancing and receding contact angles), separately. Measured values of these quantities are plotted and shown in Figure 5.a and listed in Table 1. At the same time, contact angle hysteresis (CAH) and a droplet of measurements were performed to study the motion and adhesion of water on the leaf's surfaces. Measured values of CAH and roll-off angles were plotted and shown in Figure 5.b It has been observed that the equilibrium CA of the Cassia tora was 152.87º ±1.06º. The presence of hierarchical structure observed on these plant surfaces well explains such higher value of contact angle. However, the CAH and the roll of angle on this plant leaf surface were respectively 26.81º ±4.81º and 19.17º ±6.23º, indicating strong interaction and the high contact area between water and solid phase. Such observation can be attributed to the surface topography of Cassia tora where large irregular convex papillae or micro-bumps (40μm - 60μm) were found to be covered sparsely with sharp edge-like nanostructures shown in Figure 3. (c-d).

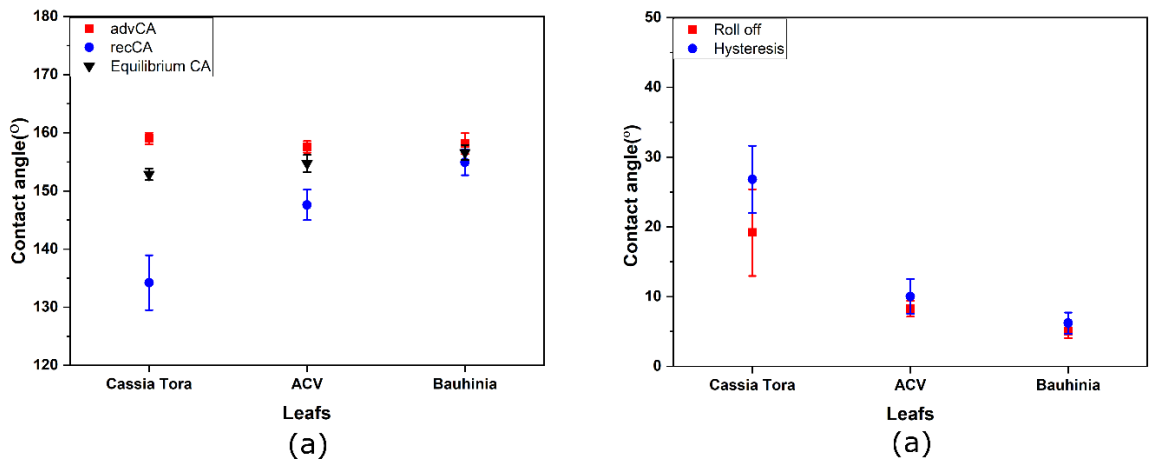

**Figure 5**. Summary of data for leaves specimens (a) Equilibrium contact angle, advancing and receding contact angle (b) Hysteresis data and roll-off angle (see also Table 1. Error bars represent the deviation for ten measurements.

Table 1. Surface wettability characteristics on different plant leaves.

| Name of the leaf | Equilibrium CA | Advancing CA | Receding CA | CA hysteresis | Roll-off angle |
|---|---|---|---|---|---|
| Cassia tora | 152.87±1.06 | 159.02±0.82 | 132.21±4.70 | 26.81±4.85 | 19.9±6.23 |
| ACV | 154.72±1.55 | 157.57±0.83 | 147.60±2.61 | 10.03±2.49 | 8.2 ±1.14 |
| Bauhinia | 156.62±1.32 | 158.15±1.68 | 151.93±2.24 | 6.28 ±1.51 | 5.0 ±0.82 |

(±) Uncertainties represent the deviation for ten measurements

Because of this sparsely spread nanostructure, the liquid(water) phase gradually replaces air entrapped within surface features when the water drop comes in contact with the surface[42]. As a result, water occupied the space between surface features, thus increasing the solid-water area with the surface. In such cases, the water drop will initially have a metastable Cassie Baxter state, which may undergo a transition into the Wenzel state, where water completely invades the texture. Thus, for such a scenario, in the beginning, the solid-water static contact angle may have a very high value, but after the transition, it will have a high value of contact angle hysteresis and roll-off angle[3,4,43,44]. For Bauhinia variegate measured equilibrium contact angle, the CAH and the roll-off angle were 156.62° ±1.3°; 6.28° ±1.5° and 5.05° ±0.82° respectively, indicating a self-cleaning superhydrophobic property of the surface. Such properties are observed when surfaces typically have closely spaced high aspect ratio nanostructures, as observed in the FESEM images of BV leaves. A stable air layer gets entrapped within such textures and acts as a cushioning layer for the water drop sitting on top of the surface. Micro ridge (with nano petals) observed on the BV leaves add further to enhance the superhydrophobic properties and, most importantly in case of dynamic wetting. The presence of these features is already being reported to significantly reduce water surface contact time, solid-liquid contact area and alter the droplet impact dynamics[45].

On the other hand, ACV shows an equilibrium CA of 154.72° ±1.55° with significantly less CAH and roll-off angle. ACV leaves have complex hierarchical morphology of papilla with closely spaced microspheres. Which are made of a three-dimensional network of fibrous structures having two tire re-entrant morphologies made up of wax. Such structures are expected to have an extremely high-water contact angle, low CAH and low roll-off angle as measured[26–28,46]. Both BV and ACV leaves are estimated to exhibit a thermodynamically stable air layer beneath water drops[8,47–52]. The stability of the droplets on the leaves is explained in terms of interfacial energy. There are two states of stability, one, when droplets breach the rough structures leading to high adhesive forces i.e., Wenzel state, and another is the Cassie Baxter state, in which the droplet rest on the top of the rough structure of the surface, trapping the air layer beneath the droplet. We first outline a thermodynamic framework that allows one to predict which of these two-state will be stable for a given water droplet, air, and substrate material.

Assuming the nanofeatures present on the leaf surface, it forms a cylindrical cavity with r radius and h height[53] (as shown in Figure 6). For state (a) (Figure 6.a), the total interfacial energy between air, water and wax is given Equation.1, and for state (b) (Figure 6.b), the total interfacial energy between air, water and wax is given by Equation.2. To have a stable air layer beneath the water droplet[54], total interfacial energy of state (b) must be less than that of state (a). (i.e., Equation.2 < Equation 1.). Further solving this will lead to $\frac{r}{2h} < 0.5198$. If the value of $\frac{r}{2h}$ for the leaf is higher than the critical value, then the water will pervade in the structure and if the value is lower than the critical value stable air layer will be formed beneath the water droplet.

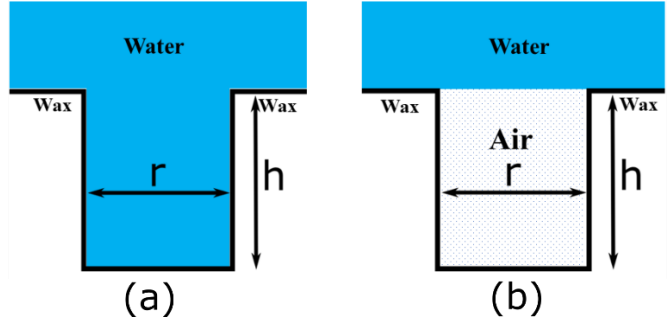

**Figure 6**. (a) Wenzel state and (b) Cassie-Baxter state formed by water droplets on the cylindrical assumed cavity with $r$ radius and $h$ height.

$$(\pi r^2 + 2\pi rh)\gamma_{wax-water} \qquad \text{Equation. 1}$$

$$(\pi r^2 \gamma_{air-water}) + (\pi r^2 \gamma_{air-wax}) + (2\pi rh\gamma_{air-wax}) \qquad \text{Equation. 2}$$

$$\therefore (\pi r^2 \gamma_{air-water}) + (\pi r^2 \gamma_{air-wax}) + (2\pi rh\gamma_{air-wax}) < (\pi r^2 + 2\pi rh)\gamma_{wax-water}$$

$$\therefore \frac{r}{2h} < \frac{(\gamma_{wax-water} - \gamma_{air-wax})}{(\gamma_{air-water} + \gamma_{air-wax} - \gamma_{air-wax})}$$

$$\therefore \frac{r}{2h} < \frac{\cos 70 \, \gamma_{air-water}}{(\gamma_{air-water} + \gamma_{air-wax} - \gamma_{air-wax})}$$

$$\therefore \frac{r(1 + \cos 110)}{2h} < \cos 70$$

$$\therefore \frac{r}{2h} < 0.5198$$

$$\therefore \frac{r}{2h}_{critical} = 0.5198$$

The $\frac{r}{2h}$ ratio for all the leaves was calculated and shown in Table S1. The value of air contact angle on a wax surface under a water environment has been calculated by using Young's equation (See supporting information). The Cassia tora leaves have two different posts spacing nanostructures, as mentioned above, widely spaced nanofeature in the valley region (sparsely spread nanofeatures) and nanostructures are close to each in the top hilly region as shown in Figure

S2. For the sparsely spread nanofeatures in cassia tora, the $\frac{r}{2h_{critical}} < \frac{r}{2h_{Cassiatora(vally)}}$ i.e., water will get inside the nanostructure creating the Wenzel state. Whereas on the top part of the Cassia tora, where the nanostructures are closely packed $\frac{r}{2h_{critical}} > \frac{r}{2h_{Cassiatora(top)}}$, creating the Cassie Baxter state.

The uniting effect of both the valley and top region of the Cassia tora leaf (shown in Figure S3), generated the metastable Cassie Baxter state, the reason for high CAH as observed experimentally. ACV and BV both have the $\frac{r}{2h_{critical}} > \frac{r}{2h}$ respectively, stating that both have thermodynamically stable air layer. The Presence of such an air layer provides superhydrophobic and high mobility self-cleaning (low CAH) properties of these surfaces as found experimentally.

### 3.3 Droplet Impact Study

Specimens of the Cassia tora, Adiantum capillus-veneris (ACV), and Bauhinia variegate were plugged just before droplet impact experiments. Drop impact conditions are conventionally specified: $\mu$ = 1.003 mPa s, $\rho$ = 998 kg/m³ and $\gamma$ = 72.7 mN m$^{-1}$ are the viscosity, density and surface tension of water in the air at 24$^0$C respectively[55]. Generally, a water droplet freely descends, collides with a superhydrophobic surface, and spreads until it attains its largest diameter. Then the droplet quickly retreats and rebounds off the surface after reaching its maximum spreading width, with and without the creation of the smaller secondary droplets.

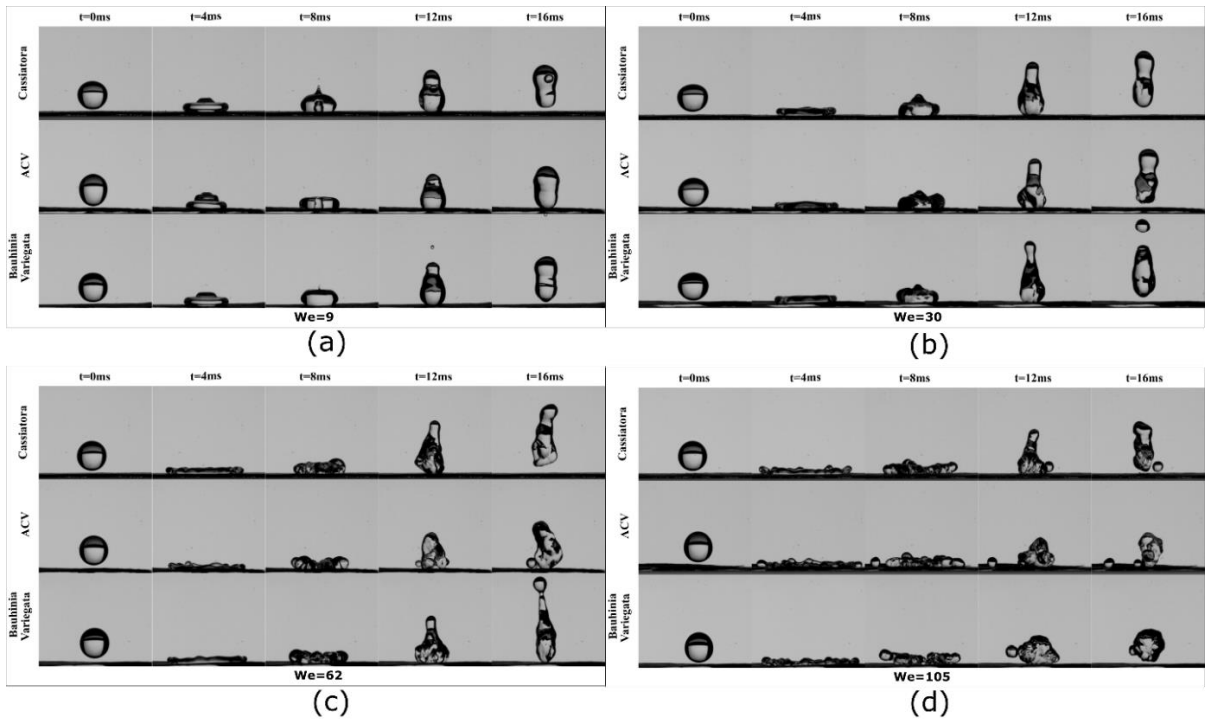

**Figure 7.** Images of droplets hitting leaves in high-speed videos, viewed from the side.

When the maximum spread diameter is attained, the droplet quickly retreats and bounces off the surface in the low We regime, as shown in Figure 7. (a) for all the leaves. When the droplet retracts, a trapped air bubble is commonly seen in this regime. The reason for air bubble entrapment is variations in the droplet's top and bottom portion's receding velocities[22]. The trapped air bubble is released or ruptured before the droplet springs off the surface. It has been observed that a micro jet discharged from the main droplet at a very high velocity often forms after the vanishing of this air bubble, though this is not always in all cases. Qualitatively, the drop impact test results agree with the behaviour observed on the natural and synthetic superhydrophobic surfaces[22,23,56,57]. Miniature secondary droplet formation is also observed in Bauhinia variegate leaves at lower *We* regimes, shown in Figure 7.(a).

In the medium *We* regime, for 30 to 62, shown in Figure 7.(b&c). The droplet bounces off the leaf surface without being pinned to the surface for all the leaves as soon as the maximum spread diameter is attained. For Bauhinia variegate leaves, a Worthington jet can be seen near the conclusion of the rebound phase in this regime; an impacting water droplet with kinetic energy takes the shape of a disc. The resistance offered to the drop by the surface is less; thus, the local retraction force is more induced because of the surface tension. The less resistance to retraction is the reason for secondary droplet formation due to the higher aspect ratio, and closely spaced nano textures and ridge structures in Bauhinia variegate. For We = 105, the droplet spreads across the surface with a higher radial momentum for all the leaves; this is for a higher We regime. Such an increased radial velocity causes the maximum spreading diameter to expand rapidly and create numerous secondary droplets. Consequently, the last primary droplet bounces off the surface without being attached to the surface shown in Figure 7. (d). Prompt splash will always occur at higher *We,* and secondary droplets always form immediately following the maximum spread diameter for all leaves [22,23,56,57].

At higher *We,* during the maximum droplet spreading in Bauhinia variegate, it has been observed that secondary droplet formation at the droplet's circumference is less when compared to ACV as shown in Figure 7.(d). The probable reason behind it is that in Bauhinia variegate, during the droplet spreading, the solid water contact area increases, restricting the spreading of droplet flow. But during retraction of the droplet, the solid water contact area decreases due to a very small-time scale, i.e., faster retraction. Thus, water will experience less restriction during the retraction phase. While in ACV, the spreading and retraction speed of the droplet doesn't have a significant difference.

During the droplet impact, it was observed that the droplet was first deformed to the extent of its maximum spreading. The droplet retreated after that and then rebounded off the surface. At the highest *We*, the condition resulted in a larger maximum spreading diameter. A wavy shape of the spreading perimeter was also produced by the droplet when it had enough kinetic energy. During the spreading phase, a droplet with enough kinetic energy created a disc shape on the solid surface. This disk's edge had a thicker rim structure than the disc itself. This thick rim had a local retraction force brought on by the surface tension, giving it a wave-like appearance at the spreading perimeter. The dual-scale hierarchical structure surface encouraged the droplet's wavy shape and fractured behaviour, as shown in Figure 7.(d), by triggering an instability of the liquid-solid interface at the disk edge and creating a secondary droplet at the rim during its maximum spreading. This splashing behaviour on the hierarchically structured surface was explored in prior work[9,16,57–62,62].

**3.4 Maximum Spreading Evaluation**

It has been extensively investigated how fluid characteristics affect the maximum spreading width of an impacting droplet[63–65], the shape of the drop[66], the surface properties[67–70] the neighbouring air pressure[71], the surface temperature,[72–77], the slope of the surface[78] or the effect of an external field[72,79–81]. Its non-dimensional form considers the maximum spreading ratio, $\beta_{max} = \frac{D_{max}}{D}$ with $D$ the initial diameter and $D_{max}$ the maximal spreading diameter of the droplet. The primary governing parameters are the Weber number i.e. $We = \frac{\rho D V_i^2}{\sigma}$ and the Reynolds number i.e $Re = \frac{\rho D V_i}{\mu}$, Figure 8. shows '$D_t$' spreading diameter of a droplet at a particular time ($T_s$) with respect to the initial diameter $D$ for all three leaves at different *We*. The kinetic, viscous, and capillary energy interaction can be used to explain the spreading and retraction processes.

The kinetic energy will drop, and the surface energy will rise during the spreading phase, while a portion of the kinetic energy will be used to counteract the viscous dissipation. The surface energy will be at its highest, and the kinetic energy will tend to zero at the end of the widening course. The surface energy will progressively become kinetic energy during the retraction operations, and some of the surface energy will also be used to combat viscous dissipation and contact line spreading. It is evident that, with the greater $We$, as $\frac{D_t}{D}$ is higher, which can be attributed to the contact line spreading more quickly due to a larger starting kinetic energy. Higher slopes of the lines correspond to a higher $We$ are indicative of faster retraction of the contact line, which is controlled by the stored surface energy at the end of the retraction phase. At the maximum spreading diameter in the capillary regime, the initial kinetic energy is transformed to surface energy. $\rho D^3 V_i^2 \sim \sigma D^2{}_{max}$ from this, we obtain $\beta_{max} \sim We^{\frac{1}{2}}$ indicating a scaling balance relating surface tension and inertia.

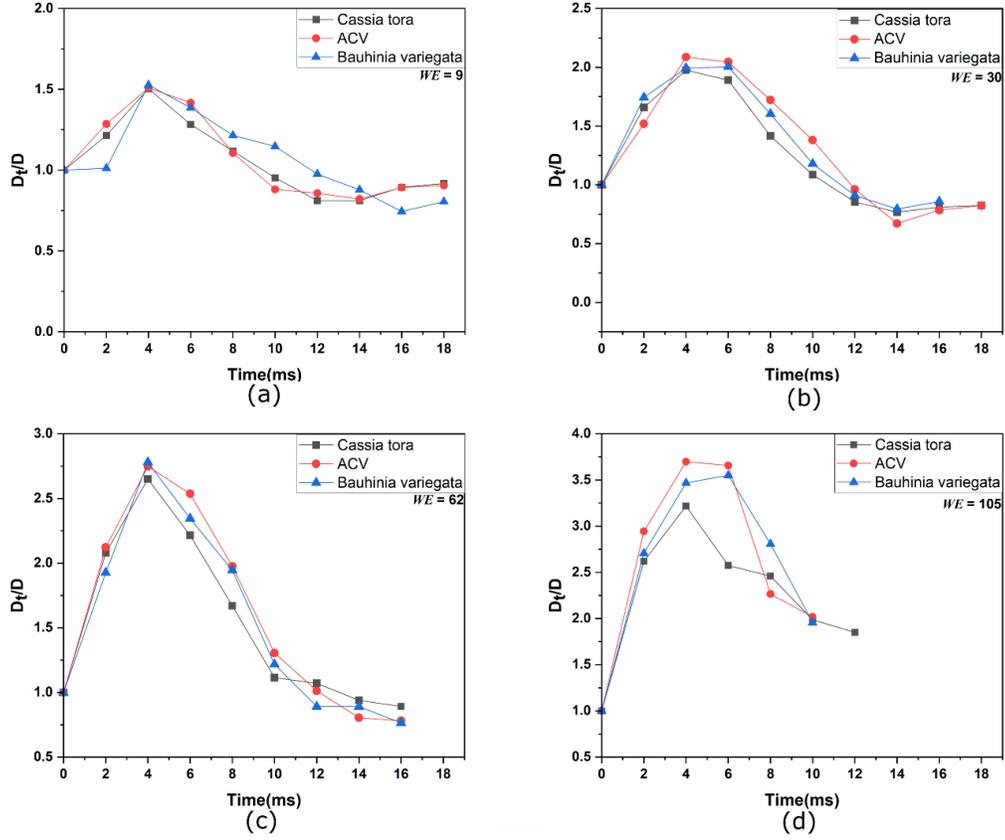

**Figure 8.** Time development of the diameters of the hitting droplet lamellas for the three leaf surfaces at various $We$ numbers.

Similarly, Up until the point of maximum spreading, the initial kinetic energy in the viscous regime is lost due to viscosity., which is the balance between the droplet's initial kinetic energy and viscous influence $\rho D^3 V_i^2 \sim \mu \left(\frac{V_i}{h}\right) D^3{}_{max}$ where $h$ is the thickness of the expanding layer. Simultaneously with volume conservation $hD_{max}^2 \sim D^3$ and, in terms of Weber number, viscous scaling can be written a $\beta_{max} \sim We^{\frac{1}{10}} Oh^{\frac{-1}{5}}$, from this, we can obtain $\beta_{max} \sim Re^{\frac{1}{5}}$. The plot between $\beta_{max}$ and $We$ the slope of the data point at higher $We$ regime matches the previous studies reported[59,61,62,82,83], from scaling law which represents at higher Weber number, inertia, and surface tension balance each other, as shown in Figure 9.(a). At lower $We$, increasing lamella dynamics were independent of the surface, as expected [59,61,62,82,83].

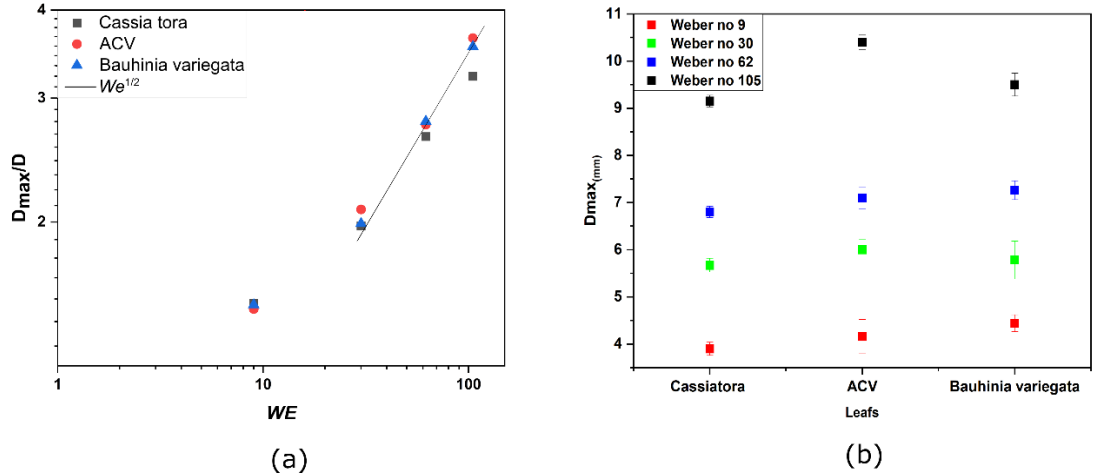

**Figure 9.** (a) Normalised maximum spread diameter, $D_{max}/D$, as a function of *We* for water. (b) $D_{max}$ for different *We* for different leaves, respectively.

Figure 9.(b) shows the plot of $D_{max}$ for different *We* for different leaves, respectively. From the plot, it has been observed that at lower *We*, the spreading diameter is maximum for BV compared to other leaves; this is because the presence of nanotextured ridges provides the minimal effective contact area of liquid and solid compared to two other surfaces; hence spreading is more in BV at a lower *We*. But as the *We* increases, this interface deforms more, and liquid touches the nanotextured valley, resulting in a higher solid-liquid contact area reflected by the lower $D_{max}$ values than ACV. On the other hand, re-entrant morphology and hierarchical structure do not allow further liquid-solid contact area even with an increase in the *We*.

## 4. Conclusions

This experiment examined the surface morphologies and the static and dynamic wettability of the three different leaves. Surface morphologies on these surfaces significantly influence the static and dynamic wettability of the water on these surfaces. The presence of sparsely spaced low aspect ratio nano textures present on Cassia tora leaf can explain high hysteresis and roll-off angle on its surface because of the slow conversion from Cassie-Baxter to Wenzel state. At lower *We,* the water-air interface has shallow surface contact on BV due to nanotextured ridge structures. Hence maximum spreading was observed on BV at lower *We*. At the higher *We*, this interface deformed more and touched the nanotextured valley resulting higher solid contact area reflected by the lower values of $D_{max}$ compared to ACV, where the presence of re-entrant morphology along with hierarchical structure doesn't allow a significant increase in water solid contact area even at high Weber number. In the vast range of *We*, droplet pinning on any leaf surface was not seen during the experiment.

**The figure for the graphical abstract/Table of content**

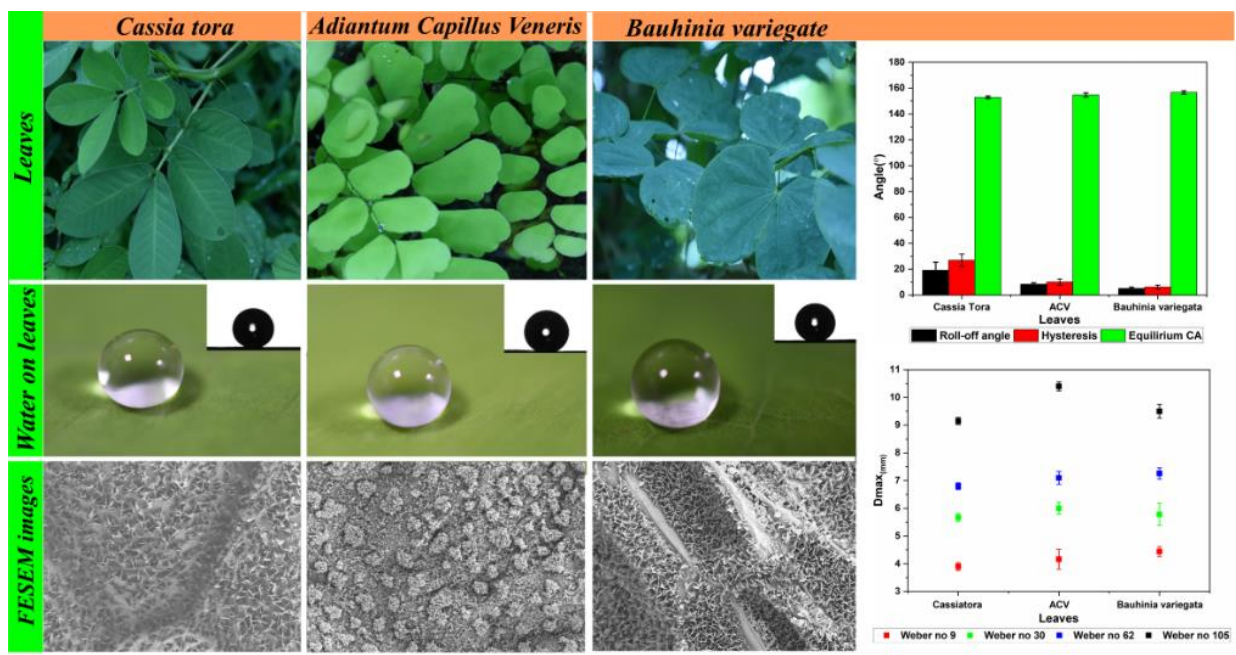

**AUTHOR INFORMATION**

**Corresponding Author**

Arindam Das*, Assistant Professor, School of mechanical sciences, Indian Institute of Technology (IIT) Goa, Email: arindam@iitgoa.ac.in, Phone: +918777747613

**Authors**

1}Shubham S. Ganar, PhD, School of School of mechanical sciences, Indian Institute of Technology (IIT) Goa, Email: shubham19263206@iitgoa.ac.in Phone: +919730083962.

2}Arindam Das*, Assistant Professor, School of mechanical sciences, Indian Institute of Technology (IIT) Goa, Email: arindam@iitgoa.ac.in, Phone: +918777747613

**Author Contributions**

The manuscript was written through the contributions of all authors. All authors have approved the final version of the manuscript.

**ACKNOWLEDGMENT**

The authors want to thank Mahesh Jarugulla for his important contribution to this work. We are also grateful to the School of Mechanical Sciences and Centre of Excellence in Particulates, Colloids and Interfaces, Indian Institute of Technology Goa, for providing the experimental facility and necessary support to conduct the above work.

**ABBREVIATIONS**

ACV, Adiantum capillus-veneris; BV, Bauhinia variegata; FESEM, Field Emission Scanning Electron Microscopy; CA, Contact angle; SHSs, Superhydrophobic self-cleaning surfaces; CAH: contact angle hyrertiesis; TPCL, Three-phase contact line; ICP, Irregular crenate platelets; $\theta$, Young contact angle; $r$, Radius of the assumed cylindrical cavity; $h$, Height of the assumed cylindrical cavity; $\gamma_{wax-water}$, Wax-Water interfacial energy; $\gamma_{air-wax}$, Air-Wax interfacial energy; $\gamma_{air-water}$, Air-Water interfacial energy; $We$, Weber number; $Re$, Reynolds number; $\beta_{max}$, The maximum spreading ratio; $D_{max}$, The maximal spreading diameter of the droplet; $D$, Initial diameter of droplet before impact; $\mu$, Viscosity; $\rho$, Fluid density; $\sigma$, Surface tension; $V_i$, The impact velocity; $D_t$, Spreading diameter of a droplet at a particular time; $T_s$, Time (sec); $t$, Thickness of the spreading droplet, $Oh$, Ohnesorge number